\documentclass[letterpaper,english,reprint, aps]{revtex4-1}
\usepackage[T1]{fontenc}
\usepackage{color}
\usepackage{babel}
\usepackage{bm}
\usepackage{amsmath}
\usepackage{amssymb}
\usepackage{graphicx}
\usepackage[unicode=true]
 {hyperref}

\makeatletter

\pdfpageheight\paperheight
\pdfpagewidth\paperwidth

\makeatother

\begin{document}
\preprint{APS/123-QED}
\title{Group Theory on Quasi-symmetry and Protected Near Degeneracy}
\author{Jiayu Li$^{1}$, Ao Zhang$^{1}$, Yuntian Liu$^{1}$, and \href{}{Qihang Liu}$^{1,2,3}$}
\email{liuqh@sustech.edu.cn}

\affiliation{$^{1}$ Department of Physics and Shenzhen Institute for Quantum Science
and Engineering (SIQSE), Southern University of Science and Technology,
Shenzhen 518055, China}
\affiliation{$^{2}$ Guangdong Provincial Key Laboratory for Computational Science
and Material Design, Southern University of Science and Technology,
Shenzhen 518055, China}
\affiliation{$^{3}$ Shenzhen Key Laboratory of Advanced Quantum Functional Materials
and Devices, Southern University of Science and Technology, Shenzhen
518055, China }
\date{\today}
\begin{abstract}
In solid state systems, group representation theory is powerful in
characterizing the behavior of quasiparticles, notably the energy
degeneracy. While conventional group theory is effective in answering
\textit{yes-or-no} questions related to symmetry breaking, its application
to determining the magnitude of energy splitting resulting from symmetry
lowering is limited. Here, we propose a theory on quasi-symmetry and
near degeneracy, thereby expanding the applicability of group theory
to address questions regarding \textit{large-or-small} energy splitting.
Defined within the degenerate subspace of an unperturbed Hamiltonian,
quasi-symmetries form an enlarged symmetry group eliminating the first-order
splitting. This framework ensures that the magnitude of splitting
arises as a second-order effect of symmetry-lowering perturbations,
such as external fields and spin-orbit coupling. We systematically
tabulate the quasi-symmetry groups within 32 crystallographic point
groups and demonstrate that $Z_{n}$, $U(1)$, and $D_{\infty}$ can
serve as the quotient group of unitary quasi-symmetry group in doubly
degenerate subspace. Applying our theory to the realistic material
AgLa, we predict a ``quasi-Dirac semimetal'' phase characterized
by two tiny-gap band anticrossings. 
\end{abstract}
\maketitle

\begin{figure}
\begin{centering}
\includegraphics[width=0.8\columnwidth]{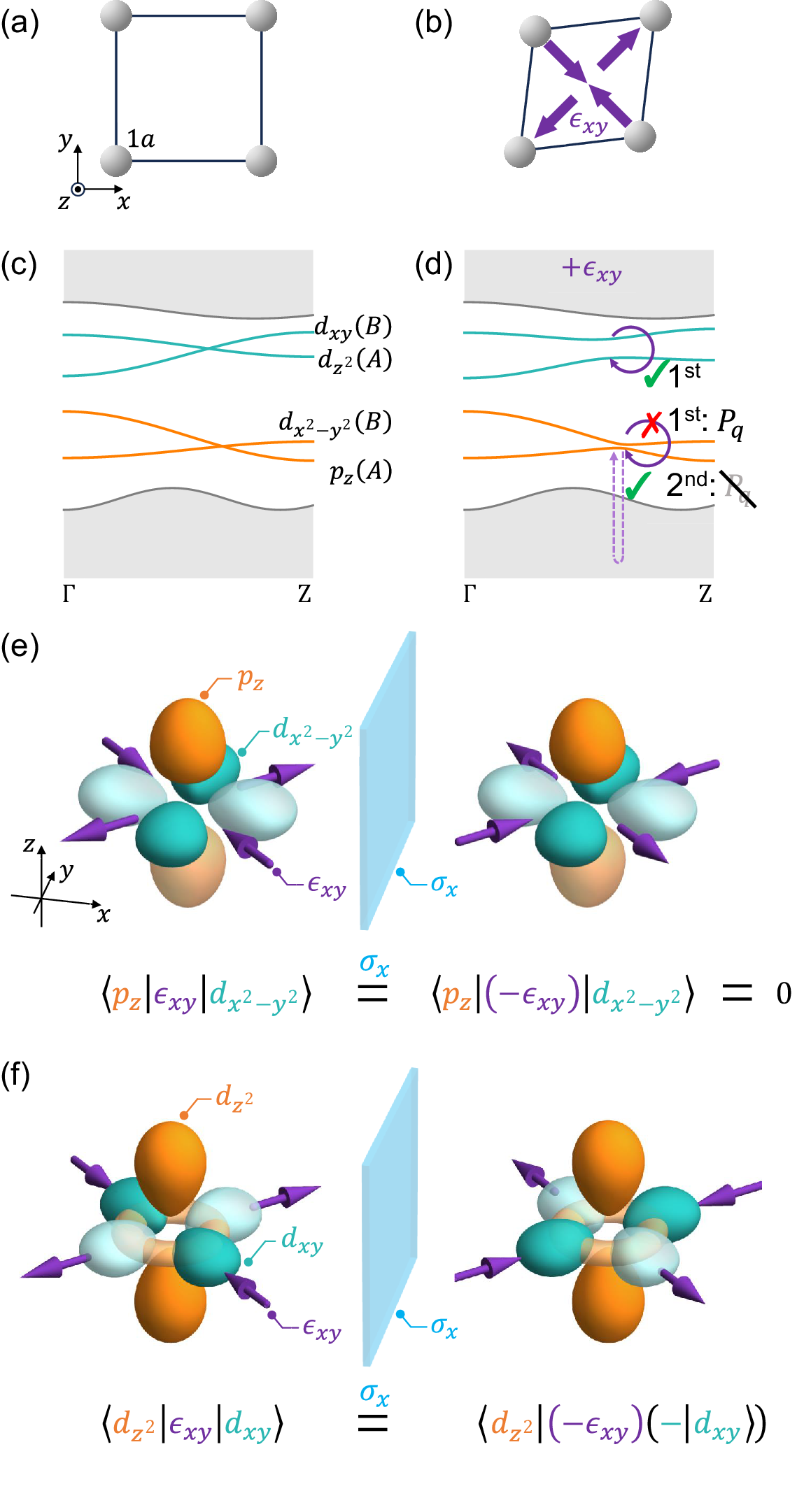}
\par\end{centering}
\caption{\label{fig:1}Schematic of a tetragonal lattice model with space group
$P4$. Top view of the tetragonal lattice without (a) and with (b)
strain $\epsilon_{xy}$. (c) Accidental band degeneracies along $\Gamma(0,0,0)$-$Z(0,0,1/2)$
line formed by $(p_{z},d_{x^{2}-y^{2}})$ and $(d_{\ensuremath{z^{2}}},d_{xy})$
orbitals settled at $1a$ Wyckoff position, where the corresponding
irreps are labelled in parentheses. (d) Strain $\epsilon_{xy}$ gaps
out both the degeneracies. Protected by quasi-symmetry $P_{q}$, the
gap opened in $(p_{z},d_{x^{2}-y^{2}})$ bands is a second-order perturbation
effect whose size is one order smaller that of $(d_{\ensuremath{z^{2}}},d_{xy})$
gap, which is a first-order effect. (e) Quasi-mirror symmetry $\sigma_{x}$
serves as the quasi-symmetry to eliminate the first-order effect in
$(p_{z},d_{x^{2}-y^{2}})$ bands. \textcolor{black}{(f) $\sigma_{x}$
would not eliminate the first-order effect in $\left(d_{z^{2}},d_{xy}\right)$
bands.}}
\end{figure}

\textit{Introduction.}---Symmetry, formulated by group theory, serves
as the most basic concept in physics, as it governs the transformation
behaviors of wavefunctions such as selection rules, conserved invariants,
and geometric phases. In solid state systems, the strength of group
representation theory applies to the behaviors of quasiparticles,
where the degeneracy of energy bands is determined by the dimension
of the irreducible representations (irreps) of little groups at certain
momenta in Brillouin zone \citep{DresselhausS2007Group,BradleyO2010TheMath}.
The recent prosperities of the field of topological phases and topological
materials, including exotic quasiparticles \citep{WanPRB2011Topological,WangPRB2012DiracSemimetal,YoungPRB2012Dirac,WangPRB2013Three,WengPRX2015Weyl,FangPRB2015Topological,LiangPRB2016Node,ArmitageRMP2018Weyl,WuPRB2018Nodal,GuoPRL2021Eight,YangPRB2021Symmetry,YuSciBull2022Encyclopedia,LiuPRB2022Systematic,ZhangPRB2022Encyclopedia,TangPRB2022Complete,TangPRL2022High,LiuPRX2022Spin,LiuTheInnovation2022Chiral,ChenArXiv2023Spin}
and novel transport responses \citep{SeemannPRB2015Symmetry,ZhangNJP2018Spin,YouPRB2018Berry,SmejkalSA2020Crystal,LiuPRL2021Intrinsic,GonzalezPRL2021Efficient,XiaoPRL2022Intrinsic,SmejkalPRX2022Giant,HuangPRL2023Intrinsic,ShaoPRL2023Neel,GonzalezArXiv2023Nonrelativistic},
are based on crystallographic groups, magnetic groups and spin groups.
It is well believed that the power of group representation theory
resides in answering the \textit{yes-or-no} questions like if the
degeneracy is lifted or if the transition matrix element is zero,
according to whether the relevant symmetry is broken. On the other
hand, the regime of group theory is hardly employed for addressing
the magnitude of energy splitting induced by symmetry lowering, because
such \textit{large-or-small} questions are supposed to related to
specific characters such as chemical environments and the strength
of perturbation. For example, consider a simple tetragonal lattice
with space group $P4$ with atomic $p_{z}$, $d_{\ensuremath{z^{2}}}$,
$d_{xy}$, and $d_{x^{2}-y^{2}}$ orbitals {[}Fig. \ref{fig:1}(a){]}.
Along the high symmetry line $\Gamma$-$Z$ with little group $C_{4}$,
the two bands, originated from $p_{z}$ (irrep $A$) and $d_{x^{2}-y^{2}}$
(irrep $B$), respectively, form an accidental degeneracy when they
meet {[}Fig. \ref{fig:1}(c){]}; so is the situation for $d_{\ensuremath{z^{2}}}$
(irrep $A$) and $d_{xy}$ (irrep $B$) orbitals. Both degeneracies
are gapped once a strain $\epsilon_{xy}$ is introduced reducing the
little-group to $C_{2}$ {[}Fig. \ref{fig:1}(b){]}, as both two matrix
elements $\langle p_{z}|\epsilon_{xy}|d_{x^{2}-y^{2}}\rangle$ and
$\langle d_{\ensuremath{z^{2}}}|\epsilon_{xy}|d_{xy}\rangle$ transform
as the identity representation of $C_{4}$. However, conventional
representation theory seems have no prediction on the gap sizes of
the two band anticrossings formed by $(p_{z},d_{x^{2}-y^{2}})$ and
$(d_{\ensuremath{z^{2}}},d_{xy})$. 

Indeed, describing or even predicting the magnitude of energy splitting
becomes increasingly essential. One notable example is that the tiny
gaps along topological nodal line, caused by spin-orbit coupling (SOC)
could lead to large Berry curvature and is thus desirable for anomalous
transport phenomena \citep{KimNatMater2018Large,SeoNature2021Colossal,OkamuraNatCommun2020Giant}.
However, such tiny gaps are typically referred to numerical results
rather than a more fundamental origin of approximate symmetry. Previous
works attempted to evaluate the degree of maintenance of approximate
symmetry by introducing fuzzy sets \citep{MezeyMolPhys1990TheConcept,KohlerCMA1991AFuzzy}
or setting artificial thresholds \citep{ZorkyJMS1996pseudosymmetry,ZwartACD2008Surprises,TalitArXiv2023Quantifying},
or to distinguish distinct topological phases protected by averaged
symmetry \citep{FuPRL2012Delocalization,MaPRX2023Average}. However,
it is unsettled that how does the magnitude of symmetry-allowed splitting
relate to approximate symmetry. Recently, it was proposed that an
$U(1)$ symmetry that commutes with the lower-order $k\cdot p$ Hamiltonian
exists as a so-called quasi-symmetry, leading to near degenerate nodal-line
structure in a chiral compound CoSi \citep{GuoNatPhys2022QuasiSymmetry,HuPRB2023Hierarchy}.
As the concept was conceived in a specific material, a comprehensive
and universal symmetry description for near degeneracy (tiny energy
splitting) is required for both fundamental understanding of group
theory and realistic material design.

In this Letter, we develop a generic theory on quasi-symmetry and
near degeneracy, and thus expand the application of group representation
theory by answering the \textit{large-or-small} question. Defined
in the degenerate subspace of an unperturbed Hamiltonian $H_{0}$,
quasi-symmetries form an enlarged symmetry group eliminating the first-order
splitting of the symmetry-lowering term $H^{\prime}$. Consequently,
the magnitude of splitting is ensured to be a second-order effect
of the symmetry-lowering perturbation such as external fields and
SOC, leading to near degeneracy. We tabulate all the possible quasi-symmetry
groups within 32 crystallographic point groups, and demonstrate that
three types of symmetry groups, \textit{i.e.}, $Z_{n}$, $U(1)$,
and $D_{\infty}$ can serve as the quotient group of unitary quasi-symmetry
group in doubly degenerate subspace. In addition to the tetragonal
model where the gap size of $(p_{z},d_{x^{2}-y^{2}})$ bands is proved
to be an order smaller than that of $(d_{\ensuremath{z^{2}}},d_{xy})$
bands {[}Fig. \ref{fig:1}(d){]}, we further apply our theory to a
realistic material AgLa to predict a SOC-driven phase transition from
Dirac nodal-line semimetal to ``quasi-Dirac semimetal'' exhibiting
two tiny-gap band anticrossings. Our work paves a new avenue for designing
materials with significant Berry curvature related properties. 

\textit{Quasi-symmetry and elimination of first-order perturbation.}---Considering
an unperturbed Hamiltonian $H_{0}$, all the symmetry operators $P_{g}$
commuting it form the symmetry group $\mathcal{G}_{H_{0}}=\left\{ P_{g}|\left[P_{g},H_{0}\right]=0\right\} $.
Once two eigenstates $\left|\psi_{\alpha}\right\rangle $ and $\left|\psi_{\beta}\right\rangle $,
labelled by two inequivalent irreps $\Gamma_{\alpha}$ and $\Gamma_{\beta}$
of $\mathcal{G}_{H_{0}}$, respectively, share the same energy $E$
of $H_{0}$, they form an accidental degeneracy \citep{Note1}. Adding
a symmetry-lowering term $H^{\prime}$ (labelled by irrep $\Gamma_{p}$),
the degeneracy splits only if the matrix element $\langle\psi_{\alpha}|H^{\prime}|\psi_{\beta}\rangle$
(labelled by $\Gamma_{\alpha}^{*}\otimes\Gamma_{p}\otimes\Gamma_{\beta}$)
transforms as the identity representation of $\mathcal{G}_{H_{0}}$,
where we termed ``$\mathcal{G}_{H_{0}}$-allowed splitting'' for
brevity (see Supplementary Section S1 \citep{SM}). As exemplified
by the tetragonal lattice model shown in Fig. \ref{fig:1}, the conventional
group representation theory has no prediction on the magnitude of
$\mathcal{G}_{H_{0}}$-allowed splitting. 

Here we demonstrate that near degeneracy, \textit{i.e.}, slightly
splitting energy levels induced by $H^{\prime}$, can also be predicted
by symmetry arguments. Such a scenario is realized when the energy
splitting is $\mathcal{G}_{H_{0}}$-allowed but $\langle\psi_{\alpha}|H^{\prime}|\psi_{\beta}\rangle$,
known as the first-order perturbation, equals zero, leading to a second-order
effect of $H^{\prime}$. We next demonstrate that the vanishment of
$\langle\psi_{\alpha}|H^{\prime}|\psi_{\beta}\rangle$ is induced
by symmetries emerged in eigensubspace of $H_{0}$, $\Psi_{\alpha\beta}=\mathrm{Span}\left(\left|\psi_{\alpha}\right\rangle ,\left|\psi_{\beta}\right\rangle \right)$.
Specifically, given a $\mathcal{G}_{H_{0}}$-allowed splitting, the
vanishment of $\langle\psi_{\alpha}|H^{\prime}|\psi_{\beta}\rangle$
can be constrained by a symmetry operator $P_{q}$ satisfying

\begin{equation}
\langle\psi_{\alpha}|H^{\prime}|\psi_{\beta}\rangle\stackrel{P_{q}}{\rightarrow}e^{i\omega\left(P_{q}\right)}\langle\psi_{\alpha}|H^{\prime}|\psi_{\beta}\rangle,\ \omega\left(P_{q}\right)\ \mathrm{mod}\ 2\pi\neq0.\label{eq:1}
\end{equation}
Note that Eq. (\ref{eq:1}) implies that $P_{q}$ is $\Psi_{\alpha\beta}$-invariant,
\textit{i.e.}, preserving the eigensubspace $P_{q}\Psi_{\alpha\beta}=\Psi_{\alpha\beta}$
and $\langle\psi_{\alpha}|H^{\prime}|\psi_{\beta}\rangle=0$ (Supplementary
Section S2 \citep{SM}). Here the phase $\omega$ is generally $\alpha,\beta$
dependent. The $\Psi_{\alpha\beta}$-invariant symmetry $P_{q}$ satisfying
Eq. (\ref{eq:1}) is thus defined as the \textit{quasi-symmetry} of
eigensubspace $\Psi_{\alpha\beta}$ in $H_{0}$, rendering the $\mathcal{G}_{H_{0}}$-allowed
splitting at least a second order effect with near degeneracy. 

Two properties related to quasi-symmetry emerge. Firstly, quasi-symmetries
are excluded from $\mathcal{G}_{H_{0}}$ because for any $P_{g}\in\mathcal{G}_{H_{0}}$,
$\omega\left(P_{q}\right)\ \mathrm{mod}\ 2\pi=0$. Secondly, the $\mathcal{G}_{H_{0}}$-allowed
splitting is exactly the second-order effect of the symmetry-lowering
term $H^{\prime}$, indicating that the concept of quasi-symmetry
can only be defined for eliminating first-order perturbed Hamiltonian.
To prove this, assuming that $P_{q}$ is a quasi-symmetry eliminating
the second-order effect ($\propto\sum_{\gamma\left(\neq\alpha,\beta\right)}\langle\psi_{\alpha}|H^{\prime}|\psi_{\gamma}\rangle\langle\psi_{\gamma}|H^{\prime}|\psi_{\beta}\rangle\left(E-E_{\gamma}\right)^{-1}$),
we note that $P_{q}$ must preserves all the eigensubspaces $\Psi_{\alpha\gamma}$
and $\Psi_{\gamma\beta}$ of $H_{0}$, yielding that $P_{g}\in\mathcal{G}_{H_{0}}$.
Thus, it rules out the possibility of ``higher-order quasi-symmetry''.
The detailed proof of both properties mentioned above is provided
in Supplementary Section S3 \citep{SM}.

\begin{figure}
\begin{centering}
\includegraphics[width=1\columnwidth]{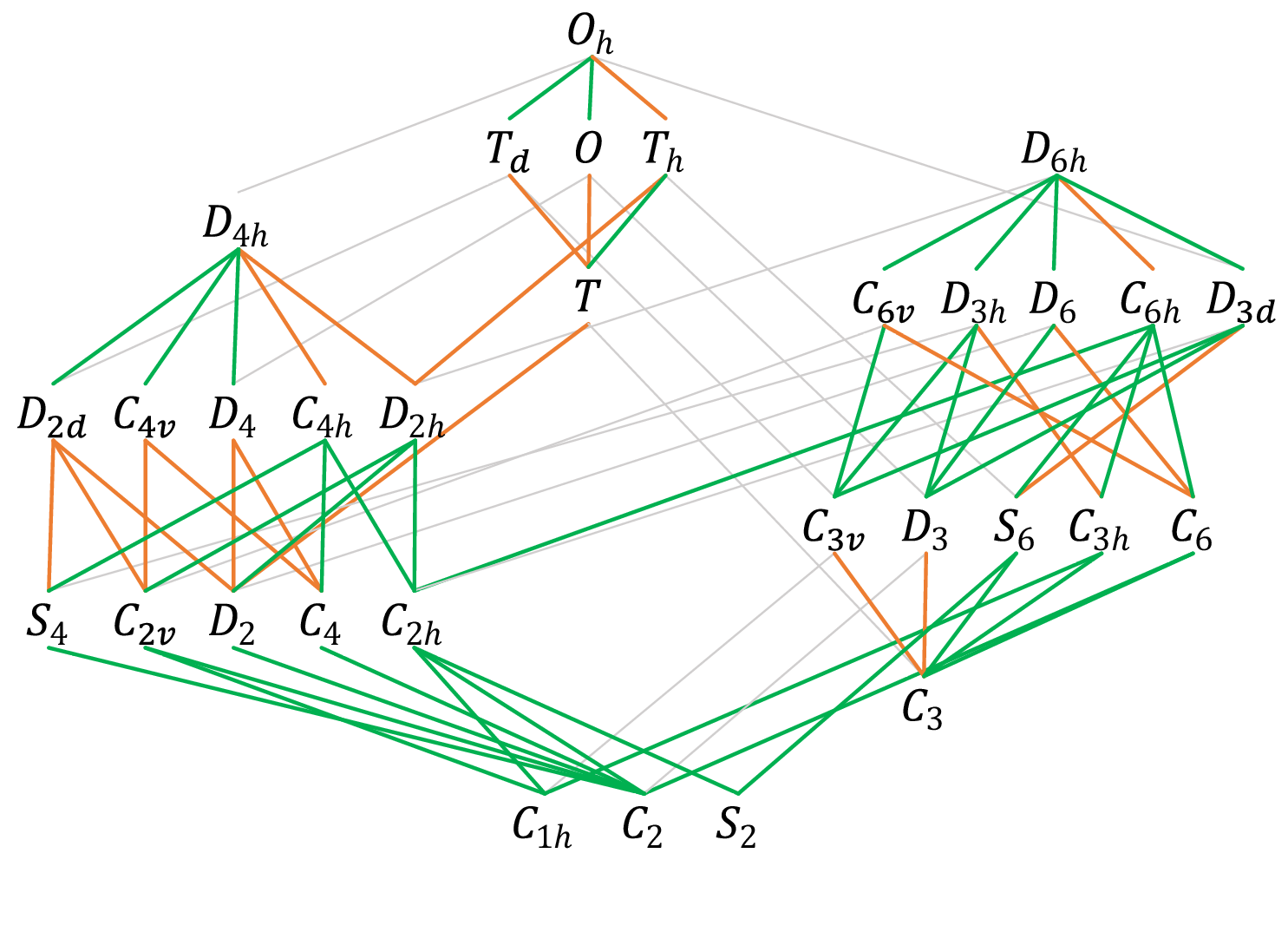}
\par\end{centering}
\caption{\label{fig:2}Quasi-symmetry group in 32 crystallographic point groups.
A group-subgroup pair is linked by green (orange) line if all (some
of) representations in the subgroup are multiple-to-one restrictive
with respect to the parent group. Hence, the parent group could serve
as the quasi-symmetry group ($\mathcal{Q}\left(\mathcal{G}_{H_{0}},P_{q}\right)$)
of the subgroup ($\mathcal{G}_{H_{0}}$). No quasi-symmetry emerges
between groups linked by gray lines. \textcolor{black}{Note that point
groups $C_{4}$ and $C_{4v}$ are involved in the tetragonal lattice
model $C_{4v}$ serves as the quasi-symmetry group of $C_{4}$, while
along the $T$ high-symmetry line of AgLa $D_{2h}$ serves as the
quasi-symmetry group of $C_{2v}$. }}
\end{figure}

\textit{Quasi-symmetry group.}---Given a symmetry group $\mathcal{G}_{H_{0}}$
and a symmetry-lowering term $H^{\prime}$, only certain eigensubspaces
$\Psi_{\alpha\beta}$ underpin quasi-symmetry. Then two crucial questions
arise: which eigensubspaces of $\mathcal{G}_{H_{0}}$ can support
quasi-symmetry, and where to look for the corresponding quasi-symmetries?
Next, we attempt to answer these by extending the exact-symmetry group
$\mathcal{G}_{H_{0}}$ to a so-called quasi-symmetry group $\mathcal{Q}\left(\mathcal{G}_{H_{0}},P_{q}\right)$,
which is hidden inside certain eigensubspace of $\mathcal{G}_{H_{0}}$.
Such extension is based on an important condition on the irreps of
quasi-symmetry $P_{q}$ that the $\mathcal{G}_{H_{0}}$-allowed splitting
$\langle\psi_{\alpha}|H^{\prime}|\psi_{\beta}\rangle$ is \textit{not}
a $\mathcal{Q}\left(\mathcal{G}_{H_{0}},P_{q}\right)$-allowed splitting.
Consequently, the matrix element $\langle\psi_{\alpha}|H^{\prime}|\psi_{\beta}\rangle$
transforms as a one-dimensional (1D) nontrivial representation of
$\mathcal{Q}\left(\mathcal{G}_{H_{0}},P_{q}\right)$, indicating that
the irreps in $\mathcal{G}_{H_{0}}$ and $\mathcal{Q}\left(\mathcal{G}_{H_{0}},P_{q}\right)$
characterizing the eigensubspace $\Psi_{\alpha\beta}$ are highly
correlated. Specifically, there must be multiple inequivalent irreps
$\Gamma_{\alpha,1}^{\prime}$ and $\Gamma_{\alpha,2}^{\prime}$ in
$\mathcal{Q}\left(\mathcal{G}_{H_{0}},P_{q}\right)$ restricting as
the same irrep $\Gamma_{\alpha}$ in $\mathcal{G}_{H_{0}}$ ($\Gamma_{\alpha,1}^{\prime}\downarrow\mathcal{G}_{H_{0}}=\Gamma_{\alpha,2}^{\prime}\downarrow\mathcal{G}_{H_{0}}=\Gamma_{\alpha}$),
termed as multiple-to-one restrictive condition (see Supplementary
Section S4 \citep{SM}).

We now tabulate all the possible eigensubspaces with all the possible
quasi-symmetry groups in crystallographic point groups. The process
is summarized in the following: 1) Starting from a point group $\mathcal{G}_{H_{0}}$
and a tentative crystallographic symmetry $P_{q}$, we construct a
quasi-symmetry group $\mathcal{Q}\left(\mathcal{G}_{H_{0}},P_{q}\right)$
by group extension as
\begin{equation}
1\rightarrow\mathcal{G}_{H_{0}}\rightarrow\mathcal{Q}\left(\mathcal{G}_{H_{0}},P_{q}\right)\rightarrow\mathcal{F}\rightarrow1,\label{eq:2}
\end{equation}
where $\mathcal{F}$ is an Abelian group generated only by $P_{q}$.
By construction, $\mathcal{G}_{H_{0}}$ is a normal subgroup of $\mathcal{Q}\left(\mathcal{G}_{H_{0}},P_{q}\right)$
and $\mathcal{Q}\left(\mathcal{G}_{H_{0}},P_{q}\right)/\mathcal{G}_{H_{0}}\cong\mathcal{F}$,
ensuring at least one irrep in $\mathcal{G}_{H_{0}}$ is multiple-to-one
restrictive (see Supplementary Section S4 \citep{SM}). 2) We tabulate
all the multiple-to-one restrictive irreps in $\mathcal{G}_{H_{0}}$.
Any eigensubspace spanned by these irreps can emerge $P_{q}$ and
other elements in $\mathcal{Q}\left(\mathcal{G}_{H_{0}},P_{q}\right)\backslash\mathcal{G}_{H_{0}}$
as quasi-symmetries.

By repeating steps 1) and 2), all the possible quasi-symmetry groups
in 32 crystallographic point groups are shown in Fig. \ref{fig:2}
and all the multiple-to-one restrictive irreps are tabulated in Tables
S1-S5 (see Supplementary Section S5 \citep{SM}). In Fig. \ref{fig:2},
a group-subgroup pair is linked by a green (orange) line if the subgroup
is normal with all (some of) irreps are multiple-to-one restrictive,
indicating that the parent group could be a quasi-symmetry group of
the subgroup. Interestingly, it is proved that for point groups, $\mathcal{Q}\left(\mathcal{G}_{H_{0}},P_{q}\right)$
can always be expressed as semidirect product $\mathcal{Q}\left(\mathcal{G}_{H_{0}},P_{q}\right)=\mathcal{G}_{H_{0}}\rtimes\mathcal{F}$
or direct product $\mathcal{G}_{H_{0}}\times\mathcal{F}$ \citep{BradleyO2010TheMath}.
Due to the completeness for crystallographic point groups, Fig. \ref{fig:2}
and Tables S1-S5 effectively facilitate the search of the quasi-symmetries
by referring to the valid group-subgroup pairs even no $P_{q}$ is
known.

It is worth noting that the constructed quasi-symmetry groups could
go beyond crystallographic point groups. Taking the double degeneracy,
the most practical case, as an example, a unitary quasi-symmetry $P_{q}$
is an element of $U(2)$. We prove that the Abelian quotient group
$\mathcal{F}$ generated by $\mathcal{Q}\left(\mathcal{G}_{H_{0}},P_{q}\right)$
must be isomorphic to three types of subgroups of $U(2)$, \textit{i.e.},
$Z_{n}$, $U(1)$ and $D_{\infty}$ (see Supplementary Section S6
\citep{SM}). For crystallographic quasi-symmetry groups shown in
Fig. \ref{fig:2}, $\mathcal{F}$ is isomorphic to $Z_{2}$ or $Z_{3}$.
On the other hand, the recent proposed near degenerate nodal line
in CoSi belongs to the case with $\mathcal{F}=U\left(1\right)$ \citep{GuoNatPhys2022QuasiSymmetry,HuPRB2023Hierarchy}.
Such a Lie group formed by quasi-symmetry is also essential for the
many-body scar dynamics \citep{RenPRL2021Quasisymmetry}. Furthermore,
the double degeneracy can also contain antiunitary quasi-symmetries,
which can be constructed by taking the complex conjugate of the eigenstates
\citep{HouFrontPhys2017Hidden}.

\textit{Tetragonal lattice model.}---To apply our theory, we now
present detailed symmetry analysis on the tetragonal lattice model
shown in Fig. \ref{fig:1}. Conventional representation theory predicts
that $\langle p_{z}|\epsilon_{xy}|d_{x^{2}-y^{2}}\rangle$ is a $C_{4}$-allowed
splitting ($\mathcal{G}_{H_{0}}=C_{4}$). We find that mirror reflection
$\sigma_{x}$, which is not in $\mathcal{G}_{H_{0}}$, is $\Psi_{p_{z},d_{x^{2}-y^{2}}}$-invariant.
Moreover, $\sigma_{x}$ reverses the strain $\epsilon_{xy}\stackrel{\sigma_{x}}{\rightarrow}-\epsilon_{xy}$
and transforms $\langle p_{z}|\epsilon_{xy}|d_{x^{2}-y^{2}}\rangle\stackrel{\sigma_{x}}{\rightarrow}-\langle p_{z}|\epsilon_{xy}|d_{x^{2}-y^{2}}\rangle$
satisfying Eq. (\ref{eq:1}) {[}Fig. \ref{fig:1}(e){]}. Therefore,
$\sigma_{x}$ is a quasi-symmetry of $\Psi_{p_{z},d_{x^{2}-y^{2}}}$
protecting the degeneracy under the first-order strain effect. Furthermore,
the quasi-reflection $\sigma_{x}$ will be broken by involving remote
states outside $\Psi_{p_{z},d_{x^{2}-y^{2}}}$, and the degeneracy
will thus be lifted by the second-order effect {[}Fig. \ref{fig:1}(d){]}
(see Supplementary Section S7 \citep{SM}). In contrast, $\sigma_{x}$
is not a quasi-symmetry of the eigensubspace spanned by $(d_{\ensuremath{z^{2}}},d_{xy})$
because $\langle d_{\ensuremath{z^{2}}}|\epsilon_{xy}|d_{xy}\rangle\stackrel{\sigma_{x}}{\rightarrow}\langle d_{\ensuremath{z^{2}}}|\epsilon_{xy}|d_{xy}\rangle$,
leading to the first-order energy splitting under $\epsilon_{xy}$,
as shown in Fig. \ref{fig:1}(f). 

The inclusion of quasi-reflection $\sigma_{x}$ expands $\mathcal{G}_{H_{0}}=C_{4}$
to the quasi-symmetry group $C_{4v}$, with the quotient group $\mathcal{F}$
isomorphic to $Z_{2}$. According to our theory, the matrix element
$\langle p_{z}|\epsilon_{xy}|d_{x^{2}-y^{2}}\rangle$ ($C_{4}$-allowed
splitting) is not a $C_{4v}$-allowed splitting, transforming as $A_{1}\otimes B_{2}\otimes B_{1}=A_{2}$,
a 1D nontrivial representation of $C_{4v}$. The irreps characterizing
$d_{x^{2}-y^{2}}$ in $C_{4v}$ and $C_{4}$ have the same dimension
$\dim B_{1}=\dim B$ during the representation restriction $B_{1}\downarrow C_{4}=B$.
Meanwhile, there is another inequivalent irreps $B_{2}$ in $C_{4v}$
restricting as the same irrep $B$ in $C_{4}$ ($B_{1}\downarrow C_{4}=B_{2}\downarrow C_{4}=B$).
In turn, by referring to Table S4 it is also straightforward to find
that $C_{4v}=C_{4}\rtimes S_{2}$ is a quasi-symmetry group of $C_{4}$,
of which irreps $A$ and $B$ support $\sigma_{x}$ as the quasi-symmetry.

\begin{figure}
\begin{centering}
\includegraphics[width=1\columnwidth]{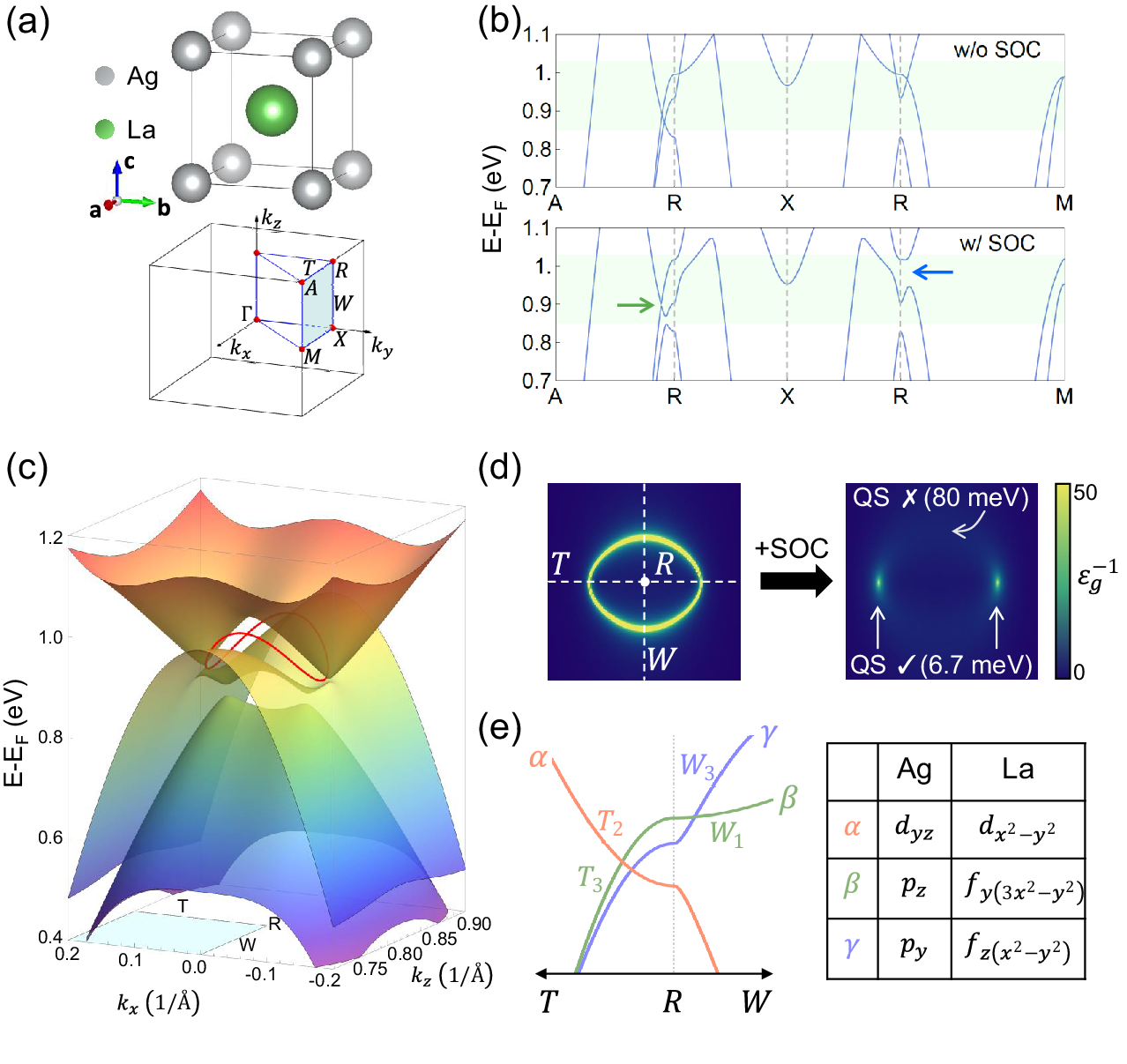}
\par\end{centering}
\caption{\label{fig:3}Quasi-symmetry protected tiny gap induced by spin-orbit
coupling in AgLa. (a) Crystal structure and Brillouin zone of AgLa.
(b) Band structure of AgLa without (upper panel) and with spin-orbit
coupling (lower panel) around the high symmetry wavevector $R$. (c)
Visualization of the quasi-Dirac semimetal phase caused by spin-orbit
coupling, where the red line marks the Dirac nodal line. (d) Inverse
gap as a function of $k_{x}$ ($T$ line) and $k_{z}$ ($W$ line)
with and without spin-orbit coupling, and \textquotedblleft QS\textquotedblright{}
denotes quasi-symmetry. (e) Orbital projections of the bands $\alpha,\beta,\gamma$
forming the nodal line along $T$ and $W$ lines, where the dominate
orbitals are tabulated. The crossing points are intersected by representations
$T_{2}$ ($W_{3}$) and $T_{3}$ ($W_{1}$) along $T$ ($W$) line.}
\end{figure}

\textit{Application to realistic material AgLa.}---We next apply
our quasi-symmetry group theory to realistic material by taking SOC
as the symmetry-lowering term. We choose AgLa (ICSD-58306 \citep{ZagoracJAC2019Recent})
as an example, which has a tetragonal structure with a space group
$P4/mmm$ and lattice constants $a=b=3.656\ \text{\AA}$ and $c=3.840\ \text{\AA}$
{[}Fig. \ref{fig:3}(a){]}. This compound has been predicted as a
topological nodal-line semimetal without SOC by topological quantum
chemistry \citep{VergnioryNature2019AComplete}. The combination of
inversion and time-reversal symmetry ensures spin degeneracy throughout
the Brillouin zone. Our calculations show that without SOC, two spin-degenerate
bands intersect around the $R$ point at $0.85\sim1\ \mathrm{eV}$
above the Fermi level {[}upper panel in Fig. \ref{fig:3}(b){]}, forming
a Dirac nodal line (DNL) on the high-symmetry plane $k_{y}=\pi/a$
{[}red curve in Fig. \ref{fig:3}(c){]}. The computational details
are shown in Supplementary Section S8 \citep{SM}. 

When SOC is considered, the entire DNL is gapped, as shown in Figs.
\ref{fig:3}(b) and \ref{fig:3}(c). However, the size of the gap
varies dramatically as a function of the wavevector. In the lower
panel of Fig. \ref{fig:3}(b) we find that the degeneracies crossed
at the $W$ ($X$-$R$) and $F$ ($R$-$M$) lines open $\sim80\ \mathrm{meV}$
gaps (blue arrow), whereas only a $6.7\ \mathrm{meV}$ gap is opened
at the $T$ ($A$-$R$) line (green arrow). We map the inverse gap
upon the $k_{x}$-$k_{z}$ plane in Fig. \ref{fig:3}(d) and observe
the minimum of the gap size at the $T$ line. Such interesting distribution
of the SOC-induced band gap reveals a novel phase transition from
DNL to quasi-Dirac semimetal, the latter of which exhibits gapless
Dirac cones under the first-order SOC effect. Such a quasi-Dirac cone
is protected by quasi-symmetry, and is slightly gapped by the second-order
SOC effect. We next show that in AgLa, the quasi-Dirac cones are located
at the $T$ line {[}Fig. \ref{fig:3}(c){]} and protected by quasi-symmetry
group $D_{2h}$.

We obtain the orbital projection of the bands relevant to the DNL
along the $T$ and $W$ lines \citep{SM}, where the $d_{yz}$, $p_{z}$,
$p_{y}$ orbitals from Ag and $d_{x^{2}-y^{2}}$, $f_{y\left(3x^{2}-y^{2}\right)}$,
$f_{z\left(x^{2}-y^{2}\right)}$ from La dominate those bands denoted
by $\alpha,\beta,\gamma$ in Fig. \ref{fig:3}(e), respectively. In
the absence of SOC, bands $\alpha$ and $\beta$ with irreps $T_{2}$
and $T_{3}$ of the little group $\mathcal{G}_{T}=C_{2v}$ intersect
with each other at the $T$ line, whereas bands $\beta$ and $\gamma$
with irreps $W_{1}$ and $W_{3}$ of the same little group $\mathcal{G}_{W}=C_{2v}$
cross at the $W$ line. We next use the theory of quasi-symmetry to
elucidate the remarkable difference of SOC-induced gap at $T$ and
$W$. By referring to Fig. \ref{fig:2} and Table S5, we find that
at wavevector $T$ only $D_{2h}$ can serves as the quasi-symmetry
group of the eigensubspace spanned by irreps ($T_{2}$,$T_{3}$) due
to the condition of multiple-to-one restrictive irreps, \textit{i.e.},
\begin{equation}
\mathcal{Q}\left(C_{2v},I\right)=D_{2h}=C_{2v}\times\left\{ E,I\right\} .\label{eq:3}
\end{equation}
Therefore, inversion $I$ emerges as the candidate of quasi-symmetry.
We further notice that by taking the on-site SOC term $H^{\prime}=H_{SOC}\propto\mathbf{L}\cdot\mathbf{S}$,
only the $z$-component $\langle\alpha,s|L_{z}S_{z}|\beta,s^{\prime}\rangle$
($s,s^{\prime}=\uparrow,\downarrow$) is $\mathcal{G}_{T}$-allowed
splitting. Such a matrix element in $D_{2h}$ transforms as $T_{2g}\otimes T_{4g}\otimes T_{3u}=T_{1u}\neq T_{1g}$,
indicating that $D_{2h}$ is indeed the quasi-symmetry group that
eliminates the first-order SOC effect. Involving the remote bands
breaks the quasi-inversion symmetry and thus opens a second-order
SOC gap of $6.7\ \mathrm{meV}$ gap. Similarly, for the $W$ line
the eigensubspace spanned by irreps ($W_{1}$,$W_{3}$) also has $D_{2h}$
as the candidate of quasi-symmetry group (see Table S5). However,
the matrix element in $D_{2h}$ transforms $\langle\beta,s|H_{SOC}|\gamma,s^{\prime}\rangle\stackrel{I}{\rightarrow}\langle\beta,s|H_{SOC}|\gamma,s^{\prime}\rangle$
with $\omega(I)\ \mod\ 2\pi=0$. Therefore, $I$ is not a quasi-symmetry
for the $W$ line according to Eq. (\ref{eq:1}), leading to a relatively
larger gap \citep{SM}. Overall, the SOC-driven quasi-Dirac semimetal
phase originates from the quasi-inversion emerged only at the $L$
line.

\textit{Discussion.}---It is worth emphasizing that our theory on
quasi-symmetry is also valid for the energy splitting of higher-dimensional
accidental degeneracy and necessary degeneracy. For the former case,
the procedure of analyzing quasi-symmetry is the same as that of doubly
degenerate band crossings. We show an example of a hexagonal lattice
model in Supplementary Section S9 \citep{SM}. In this case, the symmetry-lowering
term $H^{\prime}$ of irrep $\Gamma_{p}$ is supposed to split the
degenerate eigensubspace $\Psi_{\alpha}$ spanned by $(\psi_{\alpha,1},\psi_{\alpha,2},\cdots,\psi_{\alpha,N},)$
of $N$-dimensional irrep $\Gamma_{\alpha}$ in $\mathcal{G}_{H_{0}}$.
The matrix elements $\langle\psi_{\alpha,i}|H^{\prime}|\psi_{\alpha,j}\rangle$
($i,j=1,\cdots,N$) are $\mathcal{G}_{H_{0}}$-allowed splitting only
if $\left[\Gamma_{\alpha}\otimes\Gamma_{\alpha}\right]\otimes\Gamma_{p}$
contains the identity representation $\Gamma_{1}$ in $\mathcal{G}_{H_{0}}$,
where $\left[\Gamma_{\alpha}\otimes\Gamma_{\alpha}\right]$ denotes
the symmetric tensor product \citep{YoungPRB2012Dirac}. The identification
of quasi-symmetry is the same as that in accidental degeneracy. For
instance, consider the triplet of $\mathcal{G}_{H_{0}}=T$ spanned
by $(p_{x},p_{y},p_{z})$ (irrep $T$), where an external electric
field $\bm{\mathcal{E}}=\mathcal{E}_{z}\hat{z}$ (transform as a partner
in irrep $T$) breaks the symmetry group down to $D_{2}$ and hence
lifts the triplet due to the condition $\left[T\otimes T\right]\otimes T\rightarrow A$.
By referring Table S2 we find that irrep $T$ can have $T_{h}$ as
the quasi-symmetry group. Furthermore, the matrix element transforms
as $\langle p_{i}|d_{z}|p_{j}\rangle\stackrel{I}{\rightarrow}-\langle p_{i}|d_{z}|p_{j}\rangle$
($i,j=x,y,z$) with $d_{z}$ the electric dipole, resulting in tiny
splitting of the triplet protected by quasi-inversion symmetry.

At last, we discuss some possible scenarios and applications for the
implementation of quasi-symmetry. For example, recent angle-resolved
photoelectron spectroscopy measurements have revealed Rashba-like
spin splitting with Kramers\textquoteright{} degeneracy around certain
momenta that lack time-reversal symmetry \citep{SakamotoPRL2009Peculiar},
which can be readily explained by the theory of quasi-symmetry \citep{TaoPRB2023RashbaLike}.
More importantly, the key application of quasi-symmetry is to generate
substantial anomalous Hall effect by introducing small gaps along
the nodal lines in magnetic materials. These small gaps result in
significant Berry curvature (see Supplementary Section S10 \citep{SM}),
while the extensive distribution of nodal lines enhances the integrated
Hall conductivity \citep{KimNatMater2018Large,SeoNature2021Colossal,OkamuraNatCommun2020Giant}.
The systematic search for such materials could be accomplished through
the exploration of quasi-symmetry in magnetic nodal-line semimetals,
which have been diagnosed using magnetic topological quantum chemistry
\citep{KruthoffPRX2017Topological,XuNature2020HighThroughput,ElcoroNatCommun2021Magnetic}.
Furthermore, it is also possible to create a high-contrast anomalous
Hall device sensitive to external field, \textit{e.g.}, tiny electromagnetic
field applied may break quasi-inversion or reflection to create a
dip in Hall signal. Overall, our research paves a new avenue for expanding
the scope of group representation theory and designing materials with
large Berry curvature and anomalous transport properties. 

\textit{Note added.}---Recently, two experiments have observed the
near-quantized double quantum spin Hall effect in twisted bilayer
WSe$_{2}$ \citep{KangArXiv2024Observation,KangN2024Evidencefractionalquantum},
indicating that new symmetry indicators of the quantum spin Hall effect
are needed. We note that such a phenomenon is protected by the spin
$U\left(1\right)$ quasi-symmetry \citep{LiuArXiv2024Quantum}, which
is covered by our generic theory in this work.
\begin{acknowledgments}
We thank Xin-Zheng Li and Junwei Liu for helpful discussions. This
work was supported by National Key R\&D Program of China under Grant
No. 2020YFA0308900, National Natural Science Foundation of China under
Grant No. 12274194, Guangdong Provincial Key Laboratory for Computational
Science and Material Design under Grant No. 2019B030301001, Shenzhen
Science and Technology Program (Grant No. RCJC20221008092722009),
the Science, Technology and Innovation Commission of Shenzhen Municipality
(Grant No. ZDSYS20190902092905285) and Center for Computational Science
and Engineering of Southern University of Science and Technology.
\end{acknowledgments}

\end{document}